\documentclass[format=sigconf,natbib=false,screen]{acmart}

\def\myauthors{David Otero and Javier Parapar}
\def\mytitle{User Preference Induction with LLMs for Offline Top-N Recommendation Evaluation}

\AtBeginDocument{%
  }

\setcopyright{acmlicensed}
\copyrightyear{2026}
\acmYear{2026}
\acmDOI{XXXXXXX.XXXXXXX}

\acmISBN{978-1-4503-XXXX-X/18/06}
\usepackage{multirow}
\usepackage{soul}
\usepackage{subcaption}
\usepackage{amsmath}
\usepackage{graphicx}
\usepackage{url}
\usepackage[inline]{enumitem}
\usepackage{acronym}
\usepackage{booktabs}
\usepackage{longtable}
\usepackage{pdflscape}
\usepackage{tabularx}
\usepackage{afterpage}
\usepackage{siunitx}
\usepackage{algpseudocode, algorithm}
\usepackage{calc}
\usepackage[backend=biber,style=acmnumeric,datamodel=acmdatamodel,backref=true,backreffloats=false,sortcites]{biblatex}
\usepackage{colortbl}
\usepackage{cleveref}
\usepackage{pifont}
\usepackage{fontawesome5}
\hypersetup{
  pdfauthor={\myauthors},
  pdftitle={\mytitle},
  pdfsubject={Information Retrieval},
}

\graphicspath{{plots/}}

\DefineBibliographyStrings{american}{%
  backrefpage={cited on p.},
  backrefpages={cited on pp.}
}

\algblock{Input}{EndInput} % Create new block with start and end command
\algnotext{EndInput}  % Hide end input command
\algblock{Output}{EndOutput} % Create new block with start and end command
\algnotext{EndOutput} % Hide end output command

\definecolor{observationbg}{RGB}{255,248,220}  % Warm cream background
\definecolor{observationtext}{RGB}{139,69,19}  % Saddle brown text

\newcommand{\myparagraph}[1]{\paragraph*{\hspace*{-\parindent}\normalsize\bf#1}}

\author{David Otero}
\email{david.otero.freijeiro@udc.es}
\orcid{0000-0003-1139-0449}
\affiliation{
  \department[0]{Information Retrieval Lab, CITIC}
  \institution{Universidade da Coru\~{n}a}
  \streetaddress{Facultad de Inform\'{a}tica, Campus de Elvi\~{n}a s/n}
  \city{A Coru\~{n}a}
  \country{Spain}
  \postcode{15071}
}

\author{Javier Parapar}
\email{javier.parapar@udc.es}
\orcid{0000-0002-5997-8252}
\affiliation{
  \department[0]{Information Retrieval Lab, CITIC}
  \institution{Universidade da Coru\~{n}a}
  \streetaddress{Facultad de Inform\'{a}tica, Campus de Elvi\~{n}a s/n}
  \city{A Coru\~{n}a}
  \country{Spain}
  \postcode{15071}
}

\begin{document}

%%%%%%%%%%%%%%%%%%%%%%%%%%%%%%%%%%%%%%%%%%%%%%%%%
% Title
%%%%%%%%%%%%%%%%%%%%%%%%%%%%%%%%%%%%%%%%%%%%%%%%%
\title[\mytitle]{\mytitle}

%%%%%%%%%%%%%%%%%%%%%%%%%%%%%%%%%%%%%%%%%%%%%%%%%
% Abstract
%%%%%%%%%%%%%%%%%%%%%%%%%%%%%%%%%%%%%%%%%%%%%%%%%
\begin{abstract}

Offline evaluation is the standard methodology for comparing top-N recommender systems, yet it relies on incomplete relevance information. In most benchmark datasets, only a small subset of user--item preferences is observed, and unjudged items are commonly treated as non-relevant. This missing-as-negative assumption can bias evaluation, penalize plausible recommendations with no recorded feedback, and favour algorithms that concentrate on popular or highly exposed items. We propose an LLM-based framework to expand relevance judgements for offline recommender evaluation. Our approach uses large language models in two complementary roles. First, a preference induction stage summarizes each user's historical interactions into a textual profile that captures their tastes and interests. Second, conditioned on this profile, an LLM acts as a relevance judge for candidate recommended items that lack observed labels in the original test data. To make this process tractable and evaluation-focused, we apply judgement expansion to a pooled candidate set built from the top-ranked outputs of multiple recommenders. The resulting enriched judgements provide additional relevance evidence for previously unobserved user--item pairs, enabling ranking metrics to be computed on a more complete basis. Experimental results show that this approach is a promising strategy for improving the robustness of offline top-N evaluation and mitigating the popularity-sensitive distortions caused by sparse feedback.

\end{abstract}

\keywords{Recommender Systems Evaluation, LLM-as-a-judge, Pooling}

\maketitle

%%%%%%%%%%%%%%%%%%%%%%%%%%%%%%%%%%%%%%%%%%%%%%%%%
% Sections
%%%%%%%%%%%%%%%%%%%%%%%%%%%%%%%%%%%%%%%%%%%%%%%%%
% !TeX spellcheck = en_GB

%%%%%%%%%%%%%%%%%%%%%%%%%%%%%%%%%%%%%%%%%%%%%%%%%
% Introduction
%%%%%%%%%%%%%%%%%%%%%%%%%%%%%%%%%%%%%%%%%%%%%%%%%
\section{Introduction}
\label{sec:intro}

Recommender Systems (RS) are a central component of modern information access, helping users navigate large item spaces by selecting a small set of potentially relevant options~\cite{Ricci2022}. Because deploying and testing recommendation algorithms with real users is costly and operationally complex, \emph{offline evaluation} remains the standard first step for model development and comparison. Offline protocols are attractive because they are efficient, reproducible, and allow controlled experimentation without requiring a live system~\cite{Gunawardana2022}.

The evaluation methodology used in recommender systems has evolved substantially over time. Early work framed recommendation as a \emph{rating prediction} task, where the objective was to estimate the exact score that a user would assign to an item~\cite{Herlocker2004,Gunawardana2022}. Under this view, algorithms were typically assessed with error-based measures such as Mean Absolute Error (MAE) or Root Mean Squared Error (RMSE). Over time, however, it became clear that accurately predicting ratings does not necessarily translate into better user-facing recommendations~\cite{McNee2006}. In most applications, users are not shown predicted scores; instead, they are presented with a ranked list of items. This has shifted the focus of evaluation towards \emph{top-N recommendation}, where the quality of the ranking is more important than the accuracy of individual score estimates. As a result, ranking-oriented metrics such as Precision, Recall, AP, and nDCG have become the standard tools for offline evaluation.

Despite this transition, offline top-N evaluation in recommender systems remains fundamentally constrained by \emph{incomplete relevance information}. In classical Information Retrieval (IR), evaluation collections are built through explicit relevance assessment, often using pooling strategies to concentrate judging effort on the most promising candidates. In contrast, recommender systems usually rely on held-out interactions extracted from historical logs. This means that only a very small fraction of the user--item space is labelled. Under standard evaluation practice, items without observed feedback in the test data are commonly treated as non-relevant. This assumption is convenient, but it is also problematic: many unobserved items are not truly irrelevant, but simply unrated or never exposed to the user. Consequently, the absence of feedback is conflated with the absence of preference.

This incompleteness has important consequences for evaluation. If a recommender retrieves an item that is genuinely relevant but missing from the test set, the system receives no credit for that recommendation. The resulting metrics may therefore underestimate recommendation quality and distort comparisons across algorithms. Prior work has shown that this issue is not merely a matter of noise, but a source of systematic bias in offline evaluation. In particular, when observed interactions are concentrated on popular items, standard protocols may favour recommenders that rank head items more aggressively, while underestimating methods that retrieve relevant but less exposed alternatives. In this sense, the problem is not only sparsity in the data, but also the evaluation bias induced by treating missing feedback as negative evidence.

Pooling offers a useful point of reference in this context. In IR, pooling has long been adopted as a practical mechanism for reducing judging costs while maintaining reliable system comparisons: instead of assessing the entire collection, relevance judgements are collected only for documents retrieved by a diverse set of systems~\cite{Voorhees2001a,Losada2017a,Lu2016}. This idea is appealing for recommender evaluation as well, since it provides a structured way to focus assessment on the subset of items that are most likely to affect the comparison of algorithms~\cite{Valcarce2020}. However, pooling alone does not eliminate incompleteness. Relevant items outside the pool may remain unjudged, and the quality of the evaluation still depends on the depth and diversity of the pooled candidates~\cite{Lu2016}. Moreover, because recommender relevance is strongly personalised, extending judgements within the pool is substantially more difficult than in standard ad hoc IR~\cite{Valcarce2020}.

Beyond incompleteness, recent work has highlighted additional issues in offline evaluation, including biases induced by data collection processes, violations of the missing-at-random assumption, and the presence of data leakage in common experimental protocols~\cite{Marlin2007,Jeunen2019,Ji2021,Zhao2022}. These issues further complicate the interpretation of offline metrics, as they can distort effectiveness estimates independently of the recommendation model itself.

In this paper, we explore whether \emph{Large Language Models} (LLMs) can help address these problems by expanding the relevance information available for offline top-N evaluation. Our approach uses LLMs in two complementary roles. First, given a user's historical interactions and item metadata, an LLM induces a textual representation of that user's preferences. This \emph{preference induction} stage produces a compact profile summarising the user's tastes and interests. Second, conditioned on that induced profile, an LLM acts as a \emph{relevance judge} for candidate recommended items that do not have observed labels in the original test data. In this way, the model is not asked to infer relevance from scratch, but to assess candidate items in the context of an explicit user preference description.

To make this process computationally feasible and methodologically aligned with established evaluation practice, we apply LLM judgement to a \emph{pooled candidate set} built from the top-ranked outputs of multiple recommenders. This design mirrors the logic of pooling in IR~\cite{Voorhees2001a}: rather than attempting to judge the full user--item space, we concentrate effort on the subset of items that are most likely to influence system ranking and metric estimates. The resulting enriched qrels provide additional relevance evidence for previously unobserved user--item pairs, allowing offline metrics to be computed on a more complete basis than the original held-out interactions alone.

Our goal is not to replace human assessment with LLMs, nor to claim that LLM-generated judgements are a perfect substitute for real user feedback. Instead, we investigate whether LLM-based preference induction and relevance assessment can serve as a practical mechanism for reducing the distortions caused by incomplete feedback in offline recommender evaluation. In particular, we study whether enriching qrels in this way can make evaluation less sensitive to popularity-skewed observations and provide a fairer basis for comparing top-N recommenders~\cite{Bellogin2017,Canamares2018,Klimashevskaia2024}. The main contributions of this work are as follows:

\begin{itemize}
    \item We frame offline top-N RS evaluation as a problem of \emph{incomplete personalised relevance judgements}, highlighting how the missing-as-negative assumption can bias effectiveness estimates, favouring popularity-oriented algorithms.
    \item We propose a two-stage LLM-based framework in which a first model induces textual user preference profiles and a second model uses those profiles to assess the relevance of previously unjudged candidate items.
    \item We integrate this framework with a pooling-based evaluation design, restricting LLM assessment to candidate items retrieved by multiple recommenders in order to make judgement expansion tractable and evaluation-focused.
    \item We empirically analyse whether the resulting enriched qrels improve the robustness of offline top-N evaluation and reduce the popularity-sensitive bias associated with sparse observed feedback.
\end{itemize}

\section{Related Work}
\label{sec:rw}

\sloppy Offline evaluation in recommender systems has progressively shifted from \emph{rating prediction} to \emph{top-N ranking}. While this transition aligns evaluation with user-facing tasks, it also exposes a fundamental limitation: evaluation is conducted under highly incomplete and biased observations of user preferences. Classical work shows that optimising rating-error objectives does not necessarily improve ranking quality~\cite{Ricci2022,Gunawardana2022,McNee2006,Cremonesi2010}, yet even ranking-based evaluation remains affected by the fact that only a small fraction of the user--item space is observed.

A substantial body of work has demonstrated that recommender feedback is \emph{missing not at random} (MNAR), shaped by exposure and popularity biases, and that standard offline protocols can systematically favour algorithms that recommend already exposed or popular items~\cite{Marlin2007,Bellogin2017,Canamares2018,Schnabel2016,Joachims2017,Saito2020}. Furthermore, evaluation outcomes are highly sensitive to protocol choices such as data splitting, negative sampling, and temporal ordering~\cite{Jeunen2019,Ji2021,Rendle2020,Dacrema2019,Zhao2022}. These findings highlight that offline top-N evaluation is not only incomplete but structurally biased, motivating the need for richer and more reliable evaluation signals~\cite{Smucker2025}.

This challenge is closely related to incomplete relevance judgements in Information Retrieval (IR). Pooling has long been used to make large-scale IR evaluation tractable by focusing human assessment on candidate sets derived from multiple systems~\cite{Voorhees2001a,Voorhees2005}. Extensive work studies how pool construction affects reliability and how targeted judging strategies can reduce cost while preserving stable comparisons~\cite{Buckley2007,Carterette2006,Lu2016,Li2017,Losada2017a}. However, transferring these ideas to recommendation is non-trivial. In IR, relevance is shared across users for a given query, whereas in recommender systems it is inherently \emph{personalised} at the user--item level. Moreover, recommender evaluation typically relies on implicit feedback rather than explicit judgements, leaving the problem of missing personalised labels largely unresolved~\cite{Valcarce2018c,Valcarce2020}.

Recently, large language models (LLMs) have been explored as automatic relevance assessors in IR. Systems such as UMBRELA and benchmarks like LLMJudge demonstrate that LLMs can produce relevance labels at scale, enabling the expansion or repair of incomplete qrels~\cite{Upadhyay2024c,Rahmani2024c}. More broadly, recent work frames these approaches within the emerging \emph{LLM-as-a-Judge} paradigm, where models evaluate system outputs using contextual evidence and explicit evaluation criteria. Prior work shows that LLM-generated judgements can improve evaluation coverage by filling judgement gaps or augmenting existing test collections~\cite{Abbasiantaeb2024,Upadhyay2024a,Upadhyay2024b}. At the same time, important limitations remain, including prompt sensitivity, evidence-packaging effects, and the need for meta-evaluation through alignment with human judgements and protocol-sensitivity analyses~\cite{Takehi2024,Rahmani2025,Soboroff2024,Alaofi2024,Otero2025b}.

In recommender systems, the use of LLMs as evaluators is emerging but remains fragmented. Recent work has proposed profile-aware LLM-as-a-Judge frameworks that leverage natural-language user profiles to assess recommendation quality. For instance, LaaJ-Profile constructs interpretable user representations from interaction histories and uses them to guide pointwise and pairwise evaluation of recommendation outputs, achieving strong agreement with human judgements~\cite{Fabbri2025}. These approaches demonstrate that LLMs can approximate user-aligned evaluation signals when provided with structured preference context.

At the same time, other lines of work integrate LLMs directly into recommendation pipelines to enhance modelling capabilities, particularly in cold-start or multi-modal scenarios~\cite{Kim2024}. In parallel, the emergence of \emph{generative recommender systems} introduces new evaluation challenges that go beyond traditional ranking accuracy, including factual correctness, hallucination, and alignment with user intent~\cite{Kolb2025}. Complementary benchmarks such as RecBench+ further highlight that traditional datasets and evaluation protocols fail to capture the reasoning and interaction requirements of modern recommendation scenarios~\cite{Huang2026}.

However, despite these advances, existing LLM-based evaluation approaches in recommender systems primarily focus on assessing recommendation outputs (e.g., explanations, dialogues, or ranked lists), rather than addressing the fundamental issue of missing personalised relevance labels in offline evaluation. In particular, even profile-aware judge frameworks rely on existing candidate recommendations and do not construct new relevance judgements for unseen user--item pairs.

Our work bridges this gap by adapting LLM-based relevance assessment to the specific challenges of recommender evaluation. Rather than evaluating recommendation outputs directly, we use LLMs to construct \emph{personalised relevance judgements} for offline top-N evaluation. Specifically, we combine pooling with a two-stage LLM pipeline (preference induction followed by relevance assessment) to selectively extend qrels on evaluation-critical user--item pairs. This approach integrates insights from IR evaluation methodology with the constraints of sparse, biased feedback in recommendation, addressing a key limitation of existing offline evaluation protocols.

% !TeX spellcheck = en_GB

%%%%%%%%%%%%%%%%%%%%%%%%%%%%%%%%%%%%%%%%%%%%%%%%%
% Method
%%%%%%%%%%%%%%%%%%%%%%%%%%%%%%%%%%%%%%%%%%%%%%%%%
\section{Induced Preference Profile for LLM-based Judgements}
\label{sec:method}

We propose a practical framework to support LLM-based relevance judgement in offline top-N recommendation evaluation. This section focuses on \emph{how} we obtain LLM judgements: first by inducing user preferences, and then by conditioning a judge model on that preference evidence. The integration of these judgements into the full evaluation protocol is presented later in Section~4.

\subsection{Problem Setting}

We consider a recommendation setting with a user set $\mathcal{U}$ and an item set $\mathcal{I}$. For each user $u\in\mathcal{U}$, we have historical interaction data (ratings plus item metadata), and we also have candidate user--item pairs $(u,i)$ that require relevance judgements.

Our objective in this section is to present a general LLM-based method to judge those candidate pairs in a personalised way, using each user's historical evidence as context. The method has two stages: (i) inducing a compact preference profile from user history, and (ii) using that profile to condition the LLM judge when scoring candidate items.

\subsection{Stage 1: Induced Preference Profile (IPP)}

\begin{table*}[ht!]
    \centering
    \caption{Example of the IPP generated for the user 193999 of MovieLens 32M.}
    \footnotesize
    \label{tab:ipp-example}
    \begin{tabularx}{\textwidth}{X}
        \toprule
        \textbf{Narrative:} The user appears to have a diverse range of movie preferences, with a focus on genres such as action, adventure, sci-fi, and drama. They tend to rate highly movies from the 80s, 90s, and 2000s, with a preference for films that have a mix of action, drama, and sci-fi elements. They also seem to enjoy movies with complex storylines and themes. The user tends to rate lower movies that are overly simplistic or lack depth. They also seem to have a soft spot for movies with iconic characters, memorable quotes, and cultural significance.
        
        \textbf{Judgement instructions}: Highly Relevant (4-5 rating):
        - Action-packed sci-fi movies from the 80s, 90s, and 2000s (e.g. Star Wars, Blade Runner, The Matrix)
        - Dramas with complex storylines and themes (e.g. 2001: A Space Odyssey, The Silence of the Lambs)
        - Movies with iconic characters and memorable quotes (e.g. Toy Story, The Big Lebowski)
        - Films with a mix of action, drama, and sci-fi elements (e.g. Mission: Impossible, Minority Report)
        Moderately Relevant (2-3 rating):
        - Light-hearted comedies (e.g. Ace Ventura: Pet Detective, The Hangover)
        - Action movies with minimal depth (e.g. Die Hard: With a Vengeance, The Rock)
        - Sci-fi movies with some complexity (e.g. Interstellar, Contact)
        Not Relevant (0-1 rating):
        - Overly simplistic movies (e.g. Independence Day, Deep Impact)
        - Movies lacking depth or complexity (e.g. Transformers, The Last Airbender)
        - Films with poor storytelling or acting (e.g. The Happening, The Last Temptation of Christ) \\
        \bottomrule
    \end{tabularx}
\end{table*}

For each user $u$, we first construct an \emph{Induced Preference Profile} (IPP) from the user's historical interactions. The input to this stage is a sequence of rated items with metadata (e.g., title, genre/category, year, and rating value). Rather than directly asking an LLM to judge candidate items without context, we first ask it to produce a compact, reusable representation of user preferences. The IPP is generated in \textbf{two complementary parts}:

\begin{itemize}[leftmargin=*]
    \item \textbf{Preference narrative}: a concise natural-language summary of stable user tastes (what the user tends to like/dislike, recurring themes, preference intensity, and notable exceptions).
    \item \textbf{Judgment instructions}: explicit decision guidance for the downstream judge model (how to weigh signals, what to prioritize, and how to handle uncertainty when evidence is partial).
\end{itemize}

This decomposition is important for modularity. The narrative captures \emph{what} the user preferences look like, while the instructions specify \emph{how} those preferences should be operationalized during relevance assessment. In other words, the output of Stage~1 is designed to be directly consumable by a second LLM acting as a judge in Stage~2. An example of a generated IPP is shown in \Cref{tab:ipp-example}.

\subsection{Stage 2: LLM Relevance Judgement}

In the second stage, we use an LLM as a personalised relevance judge. For each candidate pair $(u,i)$ requiring a missing label, the model receives candidate item metadata together with user preference evidence, and outputs a discrete relevance score.

The key intuition follows directly from the motivation in Sections~\ref{sec:intro} and~\ref{sec:rw}: missing labels are not randomly distributed, and judging items without user-specific context amplifies noise and popularity priors. Conditioning the judge on user evidence anchors decisions to the user's own taste profile, making judgements more personalised and less dependent on global item popularity.

Operationally, we perform this process over candidate items selected through pooling. Thus, LLM judging effort is concentrated on evaluation-critical pairs that are most likely to affect system ranking comparisons, instead of attempting to judge the full user--item space.

To study this conditioning effect, we evaluate three strategies that progressively vary the amount and type of user evidence given to the judge:

\begin{itemize}[leftmargin=*]
    \item \textbf{IPP-only}: the judge receives only the IPP (narrative + judgement instructions), plus candidate item metadata;
    \item \textbf{Profile-only}: the judge receives only the explicit user profile (historical ratings and associated item metadata), plus candidate item metadata;
    \item \textbf{IPP + Profile}: the judge receives both IPP and explicit user profile, plus candidate item metadata.
\end{itemize}

This comparison makes the role of profile information explicit: we test whether compact preference abstraction (IPP), raw behavioural evidence (profile), or their combination (hybrid) yields the most reliable LLM judgements.

% !TeX spellcheck = en_GB

%%%%%%%%%%%%%%%%%%%%%%%%%%%%%%%%%%%%%%%%%%%%%%%%%
% Experiments
%%%%%%%%%%%%%%%%%%%%%%%%%%%%%%%%%%%%%%%%%%%%%%%%%
\section{Experiments}
\label{sec:exps}

In \Cref{sec:method}, we described how we use LLMs to produce personalised relevance judgements in recommender-system evaluation. In this section, we present the full experimental pipeline we designed to evaluate how LLM-based judgements affect offline evaluation. Our experiments are guided by four research questions:

\begin{itemize}[leftmargin=*]
    \item \textbf{RQ1:} Are LLM-based personalised judgements accurate enough to be reliable for evaluation?
    \item \textbf{RQ2:} How do these judgements affect system-ranking stability when used to fill missing labels?
    \item \textbf{RQ3:} To what extent does LLM-based hole filling reduce popularity bias in offline top-N evaluation?
    \item \textbf{RQ4:} What happens when missing labels are judged in a fully head-skewed catalogue?
\end{itemize}

The following subsections describe datasets, models, baselines, and experiments in detail. With the aim of making our experiments fully reproducible, we have released the code and all data we employed.\footnote{\url{https://github.com/IRLab-UDC/recsysipp}}

\subsection{Data, Models and Metrics}

\myparagraph{Data} We conduct experiments on two public benchmarks, MovieLens 32M\footnote{\url{https://grouplens.org/datasets/movielens/32m}} and Goodbooks-10k\footnote{\url{https://fastml.com/goodbooks-10k-a-new-dataset-for-book-recommendations}}. In \Cref{tab:datasets}, we summarise the datasets we use. Most of our main experiments are conducted on MovieLens 32M because it is larger and less head-skewed than Goodbooks-10k. We create training and test splits by assigning 80\% of each user's ratings to training and the remaining 20\% to test. The training split contains the user history used to prompt the LLM to generate the IPP.

\begin{table}[b]
    \centering
    \caption{Datasets used in the evaluation.}
    \label{tab:datasets}
    \begin{tabular}{lrrrr}
        \toprule
        \textbf{Dataset} & \textbf{Users} & \textbf{Items} & \textbf{Ratings} & \textbf{Density} \\
        \midrule
        MovieLens 32M & 200,948 & 87,585 & 32,000,204 & 0.182\% \\
        Goodbooks-10k & 53,424 & 10,000 & 5,976,479 & 1.119\% \\
        \bottomrule
    \end{tabular}
\end{table}

\myparagraph{Recommender Systems.} We select a diverse set of algorithms to obtain a representative set of recommendation techniques. To increase the diversity of pooled recommendations, some runs use the same algorithm with different parameter settings. We evaluate 31 runs for each dataset.

\begin{itemize}[leftmargin=*]
    \item \textbf{Random, Popularity}: basic baselines.
    \item \textbf{SLIM}~\cite{Ning2011}: sparse linear methods.
    \item \textbf{SVD, PureSVD, BPRMF, WRMF}~\cite{Hu2008,Cremonesi2010,Rendle2009,Takacs2009}: matrix factorization techniques.
    \item \textbf{PLSA}~\cite{Hofmann2004}: Probabilistic Latent Semantic Analysis.
    \item \textbf{NNCosNgbr-UB, NNCosNgbr-IB}~\cite{Cremonesi2010}: user-based and item-based versions of a neighbourhood technique.
    \item \textbf{LDA}~\cite{Blei2003}: Latent Dirichlet Allocation.
    \item \textbf{LRec}~\cite{Sedhain2016}: a collaborative filtering technique that employs linear logistic regression.
    \item \textbf{WSR-IB, WSR-UB}~\cite{valcarce2016language}: a simple but effective neighbourhood-based recommender, using cosine as similarity measure.
    \item \textbf{Prefs2vec}~\cite{Valcarce2019}: a method that employs cbow-based embeddings for top-N recommendation.
    \item \textbf{EASE}: a shallow item-based collaborative filtering method based on a linear autoencoder~\cite{Steck2019}.
    \item \textbf{KLD, CHI2, RSV}: neighbourhood-based techniques that stem from Rocchio’s feedback model~\cite{Valcarce2016}.
    \item \textbf{LM-WSR-UB, LM-WSR-IB}: user-based and item-based approaches that use language models for neighbourhoods~\cite{valcarce2016language}.
\end{itemize}

\myparagraph{LLM judge model}

We use a single model across the whole pipeline: \texttt{nvidia/Llama-4-Scout-17B-16E-Instruct-FP4}. The same model is used in both stages: (i) preference induction (narrative + judgement instructions) and (ii) personalised relevance judgement. All inference is executed with \texttt{vLLM}~\footnote{\url{https://vllm.ai}}. This setup provides a consistent modelling condition across experiments, so differences between results are attributable to prompting strategy and judgement context (narrative, profile, or hybrid), rather than to changes in the underlying LLM.

\myparagraph{Metrics} We use NDCG@100 and AP@100 as evaluation metrics for scoring recommender outputs.

\subsection{Experiment 0: User Subsampling}

Using an LLM to judge pooled candidates from multiple recommenders is computationally expensive in large recommendation collections, mainly because of the number of users involved. Evaluating the full pool for all users would make experimentation substantially slower than what is practically feasible.

For this reason, the first step in our experimental setup is to select a representative user subset that preserves the same evaluation behaviour as the full user set. In particular, we require the subsampled setting to remain aligned with full-user evaluation in two dimensions: (i) high rank-correlation agreement between recommender rankings, and (ii) no significant differences when comparing the same run under full vs. subsampled evaluation.

To achieve this, we design a stratified sampling procedure based on user profile length. Users are partitioned by activity levels, and we sample proportionally across strata so that the subsample preserves the profile-size distribution of the full population.

We then select a subset size large enough to satisfy both validation criteria above: high Kendall agreement with full-user rankings and zero significant run-level differences across settings.
%The pseudocode of this procedure is shown in Algorithm~\ref{alg:user-subsampling}.

% \begin{algorithm}[t]
    %     \caption{Stratified sampling of users for evaluation}
    %     \small
    %     \label{alg:user-subsampling}
    %     \begin{algorithmic}[1]
        %         \Input
        %         \Desc{$Q_H$:}{Human qrels}
        %         \Desc{$A$:}{Set of recommenders under evaluation}
        %         \Desc{$n$:}{Target sample size}
        %         \EndInput
        %         \Output
        %         \Desc{$U_{\text{sub}}$:}{User subsample}
        %         \Desc{$\tau_M, \tau_N$:}{Ranking correlations (MAP, NDCG)}
        %         \Desc{$n_{\text{equiv}}$:}{Number of equivalent runs}
        %         \EndOutput
        %         \State Compute user activity: $r_u \gets |\{(u,i,s) \in Q_{\text{human}}\}|$
        %         \State Partition users into activity quartiles and sample $n/4$ from each quartile $\to U_{\text{sub}}$
        %         \State Build subsampled qrels: $Q_{\text{sub}} \gets \{(u,i,s) \in Q_{\text{human}} : u \in U_{\text{sub}}\}$
        %         \ForAll{recommender $a \in A$}
        %         \State Evaluate $a$ on $Q_H$ and $Q_{\text{sub}}$
        %         \EndFor
        %         \State $\tau_M \gets \text{Kendall}(\text{rank}(\{m_a^H\}), \text{rank}(\{m_a^{\text{sub}}\}))$
        %         \State $\tau_N \gets \text{Kendall}(\text{rank}(\{n_a^H\}), \text{rank}(\{n_a^{\text{sub}}\}))$
        %         \State $n_{\text{equiv}} \gets |\{a \in A : \text{test}(a, Q_{\text{human}}, Q_{\text{sub}}) \text{ not significant}\}|$
        %         \State \Return $U_{\text{sub}}, \tau_M, \tau_N, n_{\text{equiv}}$
        %     \end{algorithmic}
    % \end{algorithm}

\begin{table}[t]
    \centering
    \caption{Subsampling validation summary by dataset. Kendall $\tau$ is computed between system rankings from full-qrels evaluation and subsampled-qrels evaluation.}
    \label{tab:subsampling-summary}
    \begin{tabular}{lccc}
        \toprule
        \textbf{Dataset} & \textbf{Users} & \textbf{$\tau$ AP@100} & \textbf{$\tau$ NDCG@100} \\
        \midrule
        MovieLens 32M & 500 & 0.9594 & 0.9525 \\
        Goodbooks-10k & 500 & 0.8260 & 0.9532 \\
        \bottomrule
    \end{tabular}
\end{table}

We run this algorithm for multiple candidate sizes $n \in \{500,\ldots,10000\}$, and the selected setting is summarized in~\Cref{tab:subsampling-summary}. For both datasets, a subsample of 500 users yielded strong agreement with system rankings obtained from the full user set.

In addition, we perform pairwise statistical testing across all baseline runs when comparing full-qrels and subsampled-qrels evaluations. Differences are non-significant for all runs, indicating that the reduced 500-user setting does not introduce statistically meaningful distortions in comparative system performance. Based on this validation, all subsequent experiments are conducted on the 500-user subset.

\subsection{Experiment 1: Are LLM-based personalised judgements accurate enough to be reliable for evaluation?}

The first experiment evaluates whether LLM judgements are reliable enough to be used in evaluation. The setup is intentionally simple and controlled: we ask the LLM to judge exactly the same user--item pairs that already have human ratings in the dataset (i.e., pairs in the test split). For each dataset, we create an 80/20 train--test split. The train split is used to build the user context (IPP/profile input), while the test split provides the target pairs to judge. This gives a direct comparison between LLM and human labels on matched user--item instances.

We report two complementary measures:

\begin{itemize}[leftmargin=*]
    \item \textbf{Label-level agreement (MAE).} Since LLM and human judgements are available for the same user--item pairs, we compute MAE directly between both labels. This measures how close the LLM is to human relevance at the instance level.
    \item \textbf{Evaluation-level agreement (Kendall's $\tau$).} Using the same judged pairs, we evaluate all recommenders twice: once with human test qrels and once with LLM qrels. We then compute Kendall's $\tau$ between both system rankings (for AP@100 and NDCG@100), which measures whether LLM judgements preserve the comparative ordering of systems obtained with human data.
\end{itemize}

This experiment validates whether LLM judgements are sufficiently reliable to be used as an evaluation signal before moving to hole-filling experiments over pooled candidates.

\subsection{Experiment 2: How do these judgements affect system-ranking stability when used to fill missing labels?}

This second experiment evaluates how LLM judgements affect system ranking when applied to pooled user--item pairs that are missing from the test split. In standard offline protocols, such unjudged pairs are typically treated as non-relevant by default, so we explicitly test two evaluation setups over the top-100 pooled candidates:

\myparagraph{Synthetic pool setup} All pooled user--item pairs are judged by the LLM; no human labels are used in the pool.
\myparagraph{Augmented pool setup} Human labels are kept wherever available, and the LLM is used only to judge pooled pairs that are missing from the test split.

Let $Q^{H}$ denote the human test-set qrels and $Q^{L}$ the LLM-produced qrels. For the augmented condition, we construct extended qrels as:

\begin{equation}
    Q^{E}(u,i)=
    \begin{cases}
        Q^{H}(u,i), & \text{if } (u,i) \in Q^{H}\\
        Q^{L}(u,i), & \text{if } (u,i) \notin Q^{H} \text{ and } (u,i) \in Q^{L}
    \end{cases}
\end{equation}

In words, when a pooled user--item pair has a human rating, we keep that rating, otherwise we prompt the LLM. Thus, human judgements always have precedence, and LLM labels are used only to fill holes in the pool. We then evaluate all baseline systems under both setups and report Kendall's $\tau$ rank correlation against the ranking induced by the human test-set qrels.

\subsection{Experiment 3: To what extent does LLM-based hole filling reduce popularity bias in offline top-N evaluation?}

Experiments 1 and 2 evaluate (i) how accurate LLM judgements are with respect to human labels, and (ii) how ranking agreement changes when LLM labels are introduced in pooled evaluation. However, our main objective is to test whether using LLM judgements to fill missing pooled pairs can reduce popularity bias in offline recommender evaluation.

To quantify this, we first define item popularity from training data. Let $\mathcal{D}_{\text{train}}$ denote the set of user--item interactions in the training split. Then, let $c(i)$ be the number of training ratings associated with item $i$:
\begin{equation}
    c(i) = \left|\{(u,i,r) \in \mathcal{D}_{\text{train}}\}\right|
\end{equation}
where $u$ denotes a user, $i$ an item, and $r$ a rating in the range $[0,5]$.
The function $c(i)$ is our item-level popularity signal.

Given a recommender $a \in \mathcal{A}$, user set $\mathcal{U}$, and top-$K$ list $L_a(u,K)$ for each user $u$, we define per-user recommendation popularity as the mean popularity of recommended items, and recommender-level \emph{Average Recommendation Popularity} (ARP) as the mean across users:
\begin{equation}
    \mathrm{ARP}(a)=\frac{1}{|\mathcal{U}|}\sum_{u\in \mathcal{U}}\frac{1}{|L_a(u,K)|}\sum_{i\in L_a(u,K)} c(i)
\end{equation}
ARP provides a global measure of how popularity-biased a recommender is: higher ARP means the system tends to recommend items that are more popular on average.

Using ARP, we obtain a ranking of recommenders according to their popularity orientation. In parallel, for a metric $M$ (AP@100 or NDCG@100) and qrels condition $Q$, we obtain an effectiveness ranking from the scores $M(a,Q)$.

We then compute Kendall's $\tau$ between both rankings:
\begin{equation}
    \phi(Q,M)=\tau\left(\mathrm{rank}(\{M(a,Q)\}_{a\in \mathcal{A}}),\;\mathrm{rank}(\{\mathrm{ARP}(a)\}_{a\in \mathcal{A}})\right)
\end{equation}
where $\mathcal{A}$ is the set of evaluated recommenders.

If, for an alternative qrels condition $Q'$, this correlation decreases relative to the human baseline $Q^{\mathrm{H}}$, we interpret that as reduced popularity coupling in evaluation. Therefore, the main metric in this experiment is the Kendall shift:
\begin{equation}
    \Delta\tau(Q',M)=\phi(Q',M)-\phi(Q^{\mathrm{H}},M)
\end{equation}
A negative $\Delta\tau$ indicates weaker alignment between effectiveness and popularity, consistent with reduced popularity bias under the evaluated qrels condition.

\subsection{Experiment 4: What happens when missing labels are judged in a fully head-skewed catalogue?}

After measuring popularity-bias reductions in Experiment 3, we run a final contrast experiment designed to test a setting where such reductions should not emerge. For this purpose, we use the Goodbooks-10k dataset.

This design follows directly from the structure of Goodbooks-10k. Unlike broader recommendation collections with a meaningful long tail, Goodbooks-10k is restricted to a catalogue of only 10,000 books, and these correspond precisely to the most popular and widely known items in the source collection. Therefore, the item universe available to the recommender systems, and later to the LLM judge, is already entirely concentrated on head items. This has an important consequence for our evaluation framework. In our top-N recommendation setting, the candidate items that can be recommended, pooled, and judged are only those contained in the dataset itself. Thus, in Goodbooks-10k, using LLMs to fill missing ratings does not expand relevance evidence over a heterogeneous catalogue containing both popular and niche items; it only expands judgements over items that are already popular by definition.

Under these conditions, we do not expect LLM-based hole filling to reduce the popularity bias of system rankings. On the contrary, if the judge assigns many positive labels to these missing user--item pairs, those new labels are necessarily attached to popular books. As a result, the evaluation process may add further positive evidence to the head of the popularity distribution, which can preserve or even strengthen the coupling between effectiveness and recommendation popularity.

To test this, we reuse the same setup as in Experiment 3, focusing on Goodbooks-10k. This experiment is intended as a contrast to the MovieLens 32M findings. Rather than testing whether LLM-based judgement expansion can uncover missing long-tail relevance, it examines the case where the catalogue itself prevents that effect because all candidate items belong to the popularity head.

% !TeX spellcheck = en_GB

%%%%%%%%%%%%%%%%%%%%%%%%%%%%%%%%%%%%%%%%%%%%%%%%%
% Results
%%%%%%%%%%%%%%%%%%%%%%%%%%%%%%%%%%%%%%%%%%%%%%%%%
\section{Results}
\label{sec:results}

In this section, we report and analyse the results of our experiments. We follow a progressive structure: first, we validate whether LLM judgements are reliable with respect to human labels (Experiment 1); second, we test how those judgements affect ranking stability when used over pooled candidates (Exp. 2); and third, we quantify whether this strategy reduces popularity-sensitive evaluation bias (Exps. 3 and 4).

\subsection{Experiment 1: Are LLM Judgements Reliable for Evaluation?}

\begin{table}
    \centering
    \caption{Evaluation of LLM-based relevance judgements against human ground truth on MovieLens 32M. Results compare different user modelling strategies (IPP, Profile, and both), considering the presence or absence of item plots.}
    \small
    \setlength{\tabcolsep}{3pt}
    \label{tab:exp1}
    \begin{tabular}{lcccccc}
        \toprule
        \multirow{2.3}{*}{\textbf{Judge}}
        & \multicolumn{2}{c}{\textbf{MAE}} 
        & \multicolumn{2}{c}{$\boldsymbol{\tau}$ \textbf{AP}} 
        & \multicolumn{2}{c}{$\boldsymbol{\tau}$ \textbf{NDCG}} \\
        \cmidrule(lr){2-3} \cmidrule(lr){4-5} \cmidrule(lr){6-7}
        & w/o plot & w/ plot & w/o plot & w/ plot & w/o plot & w/ plot \\
        \midrule
        
        IPP 
        & 1.1123 & 1.0300 
        & 0.9308 & 0.9282 
        & 0.9810 & 0.9810 \\
        
        Profile 
        & 0.6238 & 0.8067 
        & 0.9499 & 0.9212 
        & 0.9905 & 0.9905 \\
        
        IPP + Profile 
        & 0.9117 & 1.0045 
        & 0.9594 & 0.9499 
        & 1.0000 & 1.0000 \\
        
        \bottomrule
    \end{tabular}
\end{table}

Table~\ref{tab:exp1} reports the results of Experiment 1 on MovieLens 32M. It summarises label-level agreement (MAE) and system-level ranking agreement (Kendall's $\tau$) for both AP@100 and NDCG@100, across the three prompting strategies (IPP, Profile, and IPP+Profile), with and without plot metadata.

In this experiment, LLM judgements are generated on the same 500-user subsample used throughout the paper, and only on user--item pairs that already belong to the held-out test judgements. Therefore, the comparison is directly between human labels and LLM labels, allowing us to assess whether LLM judgements are accurate enough for evaluation use.

Overall, results show high ranking consistency with human-based evaluation, even when label-level errors are non-negligible. In MovieLens 32M, MAE is lowest for Profile (0.6238 without plot), while the strongest correlation is obtained by IPP+Profile ($\tau=0.9594$ for AP@100 and $\tau=1.0000$ for NDCG@100 without plot).

The observed MAE range is also compatible with rating noise documented in human-centric studies. Amatriain et al.~\cite{Amatriain2009a,Amatriain2009b} report that users re-rating the same items exhibit measurable inconsistency. In that context, our LLM errors are in a plausible range, while system-level agreement remains strong enough to support LLM labels as an evaluation signal.

Having established that LLM judgements preserve system-level behaviour reasonably well, we next evaluate what happens when those judgements are moved from held-out test pairs to pooled candidate pairs with missing labels.

\subsection{Experiment 2: How do these judgements affect system-ranking stability when used to fill missing labels?}

In this experiment, we focus on ranking stability under pool-based completion in MovieLens 32M. The goal is to measure how closely system rankings under LLM-based pool setups track the reference ranking obtained with human test-set qrels.

Table~\ref{tab:exp2} reports the results of Experiment 2 on MovieLens 32M. The table shows Kendall's $\tau$ rank correlation between system rankings obtained with human test-set qrels and rankings obtained under two LLM-based pool setups: \emph{Synthetic pool} (LLM labels for all pooled pairs) and \emph{Augmented pool} (human labels preserved, LLM labels only for missing pooled pairs).

For each judge setup (IPP, Profile, IPP+Profile), the table reports two ranking-agreement values per setup: one for AP@100 and one for NDCG@100. Therefore, each row summarizes how closely the evaluation under a given pool setup reproduces the reference human-based system ordering for both metrics in MovieLens 32M.

\begin{table}
    \centering
    \caption{Experiment 2 results on MovieLens 32M. Kendall's $\tau$ rank correlation between system rankings obtained with human test-set qrels and rankings obtained under LLM-based pool completion. We report synthetic pool (LLM-only judgements in the pool) and augmented pool (human labels preserved, LLM labels only for missing pooled pairs), for AP@100 and NDCG@100.}
    \small
    \label{tab:exp2}
    \begin{tabular}{lcccc}
        \toprule
        \multirow{2.3}{*}{\textbf{Judge}} & \multicolumn{2}{c}{\textbf{Synthetic pool}} & \multicolumn{2}{c}{\textbf{Augmented pool}} \\
        \cmidrule(lr){2-3} \cmidrule(lr){4-5}
        & $\boldsymbol{\tau}$ AP & $\boldsymbol{\tau}$ NDCG & $\boldsymbol{\tau}$ AP & $\boldsymbol{\tau}$ NDCG \\
        \midrule
        
        IPP 
        & 0.7714 & 0.7810 & 0.7905 & 0.8000 \\
        Profile 
        & 0.8000 & 0.7905 & 0.7714 & 0.8190 \\
        IPP + Profile 
        & 0.7837 & 0.7799 & 0.7837 & 0.7866\\
        
        \bottomrule
    \end{tabular}
\end{table}

The ranking correlations in Table~\ref{tab:exp2} are moderate rather than near-perfect, especially in the augmented pool setup. This is not necessarily a negative outcome. If pool expansion produced almost identical system rankings to the held-out human test split, there would be little practical reason to add LLM judgements in the first place. The key question is therefore not only whether rankings shift, but whether those shifts move evaluation in a less popularity-coupled direction.

This motivates Experiment 3, where we directly measure how pool completion changes effectiveness--popularity coupling through $\Delta\tau$ with ARP.

\subsection{Experiment 3: To what extent does LLM-based hole filling reduce popularity bias in offline top-N evaluation?}

This experiment addresses our central question: whether LLM-based hole filling mitigates popularity-sensitive bias in offline evaluation. We analyse this through changes in correlation between effectiveness-based system rankings and ARP-based popularity rankings.

Table~\ref{tab:exp3} reports the Experiment 3 results on MovieLens 32M. It shows $\Delta\tau$, i.e., the change in Kendall correlation between effectiveness-based system rankings (AP@100, NDCG@100) and ARP-based popularity ranking, relative to the human-labelled baseline. Rows are organised by judge setup (IPP, Profile, IPP+Profile), and columns separate \emph{Synthetic Pool} and \emph{Augmented Pool} for AP@100 and NDCG@100. Negative values indicate lower effectiveness--popularity coupling than the human baseline, while positive values indicate higher coupling.

The strongest reduction is obtained by IPP+Profile (consistently $\Delta\tau=-0.2095$ or $-0.2000$), followed by IPP. By contrast, \textbf{Profile} alone is mixed: it slightly reduces coupling for NDCG but increases coupling for AP. This suggests that exposing only raw profile evidence is less effective at controlling popularity alignment than using IPP-based preference abstraction, especially when combined with profile context.

The stronger reduction in popularity coupling observed with IPP-based judgements can be explained by the different kind of evidence exposed to the judge. Under the Profile setup, the LLM directly sees the user’s historical interactions, including explicit mentions of previously rated items and their associated scores. Because popular items appear more often in users’ histories, and because the model’s parametric knowledge has likely internalised frequent associations between well-known items and strong positive signals, this context can activate a popularity prior at judgement time. As a result, when the judge encounters a candidate item that is itself popular, or semantically close to many popular items already present in the profile, it may be more likely to assign an overly favourable relevance score. In other words, the raw profile not only provides preference evidence; it also exposes item-level popularity cues that the judge may inadvertently exploit.

By contrast, the IPP-only condition provides a compressed description of the user’s tastes rather than a list of concrete past interactions. This abstraction still conveys what the user tends to like, but it removes many of the direct in-context triggers linked to specific popular items and explicit rating patterns. Without those concrete popularity-laden anchors in the prompt, the judge is encouraged to base its decision on broader preference dimensions (such as genre, style, theme, or complexity) instead of reproducing associations tied to exposure frequency. This helps explain why IPP judgements yield a more negative $\Delta \tau$: they weaken the alignment between measured effectiveness and recommendation popularity, which is consistent with stronger mitigation of popularity bias in offline evaluation.

Taken together, the three experiments provide a coherent picture: LLM judgements are sufficiently reliable on known labels, remain stable when transferred to pooled completion settings, and can reduce popularity-coupled distortions under appropriate prompting conditions.

\begin{table}
    \centering
    \caption{Experiment 3 results on MovieLens 32M. Impact of LLM-based evaluation on popularity bias, reported as the change in correlation ($\Delta \tau$) between system rankings (AP, NDCG) and popularity (ARP), using the human-labelled pool as the baseline. Negative values indicate a reduction in popularity bias.}
    \small
    \label{tab:exp3}
    \begin{tabular}{lcccc}
        \toprule
        \multirow{2.3}{*}{\textbf{Judge}} & \multicolumn{2}{c}{\textbf{Synthetic Pool}} & \multicolumn{2}{c}{\textbf{Augmented Pool}} \\
        \cmidrule(lr){2-3} \cmidrule(lr){4-5}
        & AP & NDCG & AP & NDCG \\
        \midrule
        
        IPP 
        & -0.1333 & -0.1238 & -0.1524 & -0.1429 \\
        Profile 
        & 0.0095 & -0.0762 & 0.0381 & -0.0857 \\
        IPP + Profile 
        & -0.2095 & -0.2000 & -0.2095 & -0.2000 \\
        
        \bottomrule
    \end{tabular}
\end{table}

\subsection{Experiment 4: What happens when missing labels are judged in a fully head-skewed catalogue?}

To complete that picture, we now examine a boundary case where those gains may not hold. In this experiment, we test the same evaluation logic on Goodbooks, a catalogue strongly concentrated on head items, to assess whether the bias-mitigation trend observed in MovieLens 32M still transfers. For Goodbooks, we focus the presentation on Experiments 1 and 3, which are the most informative.

We start with judgement quality on held-out human labels (same setup as Experiment 1). Results are shown in \Cref{tab:exp1-gb}.

\begin{table}
    \centering
    \caption{Evaluation of LLM-based relevance judgements against human ground truth on Goodbooks. Results compare different user modelling strategies (IPP, Profile, and both), considering the presence or absence of item plots.}
    \small
    \setlength{\tabcolsep}{3pt}
    \label{tab:exp1-gb}
    \begin{tabular}{lcccccc}
        \toprule
        \multirow{2.3}{*}{\textbf{Judge}}
        & \multicolumn{2}{c}{\textbf{MAE}} 
        & \multicolumn{2}{c}{$\boldsymbol{\tau}$ \textbf{AP}} 
        & \multicolumn{2}{c}{$\boldsymbol{\tau}$ \textbf{NDCG}} \\
        \cmidrule(lr){2-3} \cmidrule(lr){4-5} \cmidrule(lr){6-7}
        & w/o plot & w/ plot & w/o plot & w/ plot & w/o plot & w/ plot \\
        \midrule
        
        IPP 
        & 1.3120 & 1.2279 
        & 0.8882 & 0.8529 
        & 0.9471 & 0.9384 \\
        
        Profile 
        & 0.6485 & 0.7055 
        & 0.8176 & 0.8798 
        & 0.9766 & 0.9766 \\
        
        IPP + Profile 
        & 1.0432 & 1.3164 
        & 0.8446 & 0.8680 
        & 0.9532 & 0.9649 \\
        
        \bottomrule
    \end{tabular}
\end{table}

As in MovieLens 32M, ranking-level agreement remains reasonably strong despite non-trivial MAE. Profile again yields the lowest MAE, while Kendall agreement stays high enough to treat LLM labels as a usable evaluation signal in this dataset as well.

One plausible explanation for this Goodbooks behaviour is that many items are widely known and semantically easier to ground through explicit interaction traces. In that setting, providing the raw user profile can reduce abstraction noise from narrative compression and help the judge map candidate books to concrete preference signals more directly. By contrast, narrative-only prompting may lose fine-grained cues that are useful for distinguishing moderately relevant from highly relevant items.

Finally, we analyse popularity coupling (same setup as Experiment 3). Results are shown in \Cref{tab:exp3-gb}. Here, the key finding differs from MovieLens 32M: LLM-based completion does not reduce popularity coupling; it increases it (positive $\Delta\tau$). Intuitively, in this dataset most candidate items in the unlabelled pool are already popular by construction. When we add new judgements for those previously unlabelled pairs, we are mostly adding relevance evidence on head items, not on tail items. As a consequence, effectiveness rankings become more aligned with ARP, which appears as stronger popularity bias.

In short, Goodbooks highlights an important boundary condition of the method: the impact of LLM hole filling depends on the popularity structure of the unlabelled pool. When missing pairs are themselves head-skewed, completion can amplify popularity coupling instead of mitigating it.

\begin{table}
    \centering
    \caption{Experiment 3 results on Goodbooks. Impact of LLM-based evaluation on popularity bias, reported as the change in correlation ($\Delta \tau$) between system rankings (AP, NDCG) and popularity (ARP), using the human-labelled pool as the baseline. Negative values indicate a reduction in popularity bias.}
    \small
    \label{tab:exp3-gb}
    \begin{tabular}{lcccc}
        \toprule
        \multirow{2.3}{*}{\textbf{Judge}} & \multicolumn{2}{c}{\textbf{Synthetic Pool}} & \multicolumn{2}{c}{\textbf{Augmented Pool}} \\
        \cmidrule(lr){2-3} \cmidrule(lr){4-5}
        & AP & NDCG & AP & NDCG \\
        \midrule
        IPP 
        & 0 & 0.1637 & 0.0117 & 0.1754 \\
        Profile 
        & -0.0584 & 0.1053 & -0.0585 & 0.0819 \\
        IPP + Profile 
        & -0.0585 & 0.0234 & -0.0234 & 0.0585 \\
        \bottomrule
    \end{tabular}
\end{table}

% !TeX spellcheck = en_GB

%%%%%%%%%%%%%%%%%%%%%%%%%%%%%%%%%%%%%%%%%%%%%%%%%
% Conclusions
%%%%%%%%%%%%%%%%%%%%%%%%%%%%%%%%%%%%%%%%%%%%%%%%%
\section{Conclusions}
\label{sec:conclusions}

This work started from a practical evaluation problem in top-$N$ recommendation: offline protocols rely on sparse held-out interactions, and missing user--item labels are commonly treated as non-relevant. As discussed in the introduction, this assumption can distort system comparison and amplify popularity-sensitive effects. To address this, we proposed a two-stage LLM framework in which a first model induces user preference profiles and a second model uses that evidence to judge missing pooled candidate pairs.

Across the experiments, our findings support three main conclusions. First, LLM judgements are sufficiently reliable to be useful in evaluation: although label-level agreement with human ratings is not perfect, system-level ranking agreement remains high. Second, when judgements are transferred from known test pairs to pooled completion settings, the resulting rankings remain reasonably stable, which suggests that the method can be integrated into comparative offline evaluation without collapsing evaluation behaviour. Third, and most importantly, using an LLM to judge previously unlabelled pooled pairs is an effective strategy to reduce popularity-sensitive distortions across prompting strategies. Prompting design still matters, but the central result is robust: popularity coupling decreases in general when missing pooled labels are completed with LLM judgements.

To further validate this interpretation, we also tested a setting where a reduction in popularity coupling should not occur. The Goodbooks experiment was deliberately designed to test this mechanism in a fully head-skewed catalogue, where we expected no reduction in popularity coupling. The observed results are consistent with that expectation: when most unlabelled pooled candidates are already popular, completion adds relevance evidence mainly on head items, so LLM-based hole filling does not reduce popularity bias and may even increase it.

Overall, LLM-based judgement expansion is a practical tool for reducing incompleteness in offline recommender evaluation and, in suitable catalogue conditions, for reducing popularity bias as well. Its impact depends on pool composition and prompting design, and the evidence across our experiments is coherent with that causal interpretation. An important direction for future work is to evaluate this approach when the recommenders themselves are LLM-based or LLM-assisted, where circularity effects may appear (e.g., an LLM judge systematically favouring recommendations produced by another LLM). Future work should also move towards adaptive completion policies that explicitly control popularity exposure, uncertainty-aware judging pipelines that prioritise selective human validation, and broader robustness analyses across datasets, domains, and pool-construction strategies.

%%%%%%%%%%%%%%%%%%%%%%%%%%%%%%%%%%%%%%%%%%%%%%%%%
% References
%%%%%%%%%%%%%%%%%%%%%%%%%%%%%%%%%%%%%%%%%%%%%%%%%
\printbibliography

\end{document}